\documentclass[12pt]{iopart}
\usepackage{graphics}
\begin{document}

\comment{Superconducting transition in Nb nanowires fabricated using focused ion beam}

\author{A Engel}

\address{University of Z\"{u}rich,
         Physics Institute,
         Winterthurerstr. 190, 8057 Z\"{u}rich, Switzerland}
\ead{andreas.engel@physik.uzh.ch}
\begin{abstract}

In a recent paper Tettamanzi \emph{et al} (2009 \emph{Nanotechnology} \textbf{20} 465302) describe the fabrication of superconducting Nb nanowires using a focused ion beam. They interpret their conductivity data in the framework of thermal and quantum phase slips below $T_c$. In the following we will argue that their analysis is inappropriate and incomplete, leading to contradictory results. Instead, we propose an interpretation of the data within a SN proximity model.

\end{abstract}

\submitto{\NT}

The fabrication of superconducting Nb nanowires utilising a focused Ga-ion beam is described by Tettamanzi \emph{et al} \cite{Tettamanzi09}. The fabricated nanowires had physical widths down to 70~nm and lengths up to several micrometers. Resistance vs temperature data as a function of the physical width of the nanowires have been obtained using a two-point setup. The data have been analysed assuming the occurrence of phase slip centres in effectively one-dimensional superconducting wires \cite{Bezryadin08}. While the developed nanostructuring technology is certainly interesting and could be important for the development of new superconducting devices, we consider the interpretation of the resistance data inappropriate and incomplete.

In figure 1 (c) in \cite{Tettamanzi09} a cross-section obtained with a high-resolution transmission electron microscope (HR-TEM) indicates the general morphology of the nanowires: a core region of Nb, presumably free of Ga contamination, is surrounded by a several ten nanometers thick layer contaminated with Ga. In addition to the contamination with Ga, the ion bombardment introduces structural defects in this outer layer. It is therefore justified to assume that its intrinsic properties are that of a dirty normal metal. Tettamanzi \emph{et al} have determined the Nb core width for three devices indicating a roughly linear relation between the core width and the physical width of the nanowire (figure 1 (d) in \cite{Tettamanzi09}). Linear inter- respectively extrapolation gives expected Nb core widths $w_\mathrm{Nb}$ between about 15 and 70 nm for the four devices labelled A to D in \cite{Tettamanzi09}. This is in stark contrast to the widths of these devices as they result from a phase slip analysis. Tettamanzi \emph{et al} list in table 1 best-fit values for the Nb width of about 1 nm for all four devices. Furthermore, best-fit values for the normal state resistances are one order of magnitude larger than measured values. A fact that cannot be rectified even if one takes into account that core and outer layer are connected in parallel. More severely, the resulting critical temperatures $T_c$ for the Nb core up to 14 K are unphysically high. Most importantly, the sudden drop in resistance occurring in devices B to D for temperatures around 6 to 6.5 K---not even mentioned in \cite{Tettamanzi09}---cannot be explained within their phase slip model. It is also worth mentioning that in the studies cited by Tettamanzi \emph{et al}, for example \cite{Bezryadin08}, $R$ vs $T$ data can usually be fitted by the phase-slip model over several orders of magnitude in resistance, compared to the extremely limited range in figure 3 in \cite{Tettamanzi09}.

\begin{figure}
\includegraphics{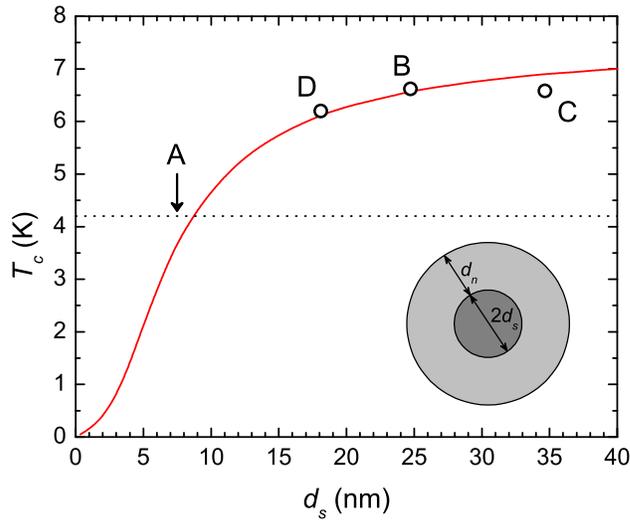}
\caption{Critical temperature $T_c$ as a function of Nb core size for devices B to D from \cite{Tettamanzi09}. The red line is the rescaled $T_c(d_s)$ data from \cite{Kushnir06}, the horizontal dashed line is the base temperature for the experiments of \cite{Tettamanzi09}. The arrow marks the expected $T_c$ for device A, which lies outside the experimentally accessible range. The inset shows a sketch of the proposed cross-section for modeling the nanowires. \label{fig.Tc}}
\end{figure}

Instead, we propose a bilayer superconducting-normal metal (SN) proximity-effect model \cite{Gennes64} that can qualitatively explain all observed features in the their resistance data, especially the second drop in resistance observed in devices B to D. It is well established that the proximity effect causes a reduction of $T_c$ on reducing the superconducting layer thickness \cite{Kushnir06} as well as in superconducting nanostructures \cite{Ilin04a}. A detailed quantitative analysis of the resistance data of \cite{Tettamanzi09} is well beyond this comment, not least because the geometry and dimensions of the nanowires require numerical approaches to determine the critical temperature $T_c$ as a function of the Nb core size. An approximate model of the nanowires' cross-section may be a cylinder with core diameter $2d_s$ and outer layer normal metal thickness $d_n$, resembling a SN bilayer with thickness $d_s$ and $d_n$, respectively, and periodic boundary conditions. Assuming that $T_c(d_s)$ approximately scales with device parameters, e.g. $d_n$ and the intrinsic $T_c$ of the superconductor, we used the numerically obtained data of \cite{Kushnir06} for a Cu/Nb/Cu trilayer and rescaled them to fit $T_c$ of devices B to D, see figure \ref{fig.Tc}. We used the data in figure 3 of \cite{Tettamanzi09} to determine $T_c$ and set $d_s=w_\mathrm{Nb}/2$, where $w_\mathrm{Nb}$ is the interpolated Nb core width as obtained from HR-TEM. Extrapolation of the rescaled $T_c(d_s)$ to infinite Nb core size gives an intrinsic $T_c$ for the core of about 7.5~K in good agreement with $T_c$ of the unstructured film. From figure \ref{fig.Tc} one can conclude that the expected $T_c$ for device A is outside the experimentally accessible range in \cite{Tettamanzi09}, and that $T_c$ in even narrower nanowires would be suppressed even more.

In conclusion Tettamanzi \emph{et al} present an interesting technology for the fabrication of proximity coupled superconducting nanowires, but by no means present any evidence for phase-slip behaviour in these nanowire devices.

\section*{References}
\bibliographystyle{unsrt}
\bibliography{Literature}

\end{document}